\documentclass[a4paper,11pt]{article}
\pdfoutput=1 

\usepackage{jheppub}
\usepackage{epsfig}
\usepackage{amsmath}
\usepackage{graphicx}
\usepackage{appendix}
\usepackage{titlesec}
\usepackage{subfigure}
\usepackage{subcaption}

\def\A0#1{\Pi_{\rm #1}(0)}
\def\AP0#1{\Pi'_{\rm #1}(0)}
\def\be{\begin{equation}}
\def\ee{\end{equation}}
\def\bea{\begin{array}}
\def\eea{\end{array}}
\def\beqa{\begin{eqnarray}}
\def\eeqa{\end{eqnarray}}
\def\beqas{\begin{eqnarray*}}
\def\eeqas{\end{eqnarray*}}

\def\bp{\begin{picture}}
\def\ep{\end{picture}}
\def\bc{\begin{center}}
\def\ec{\end{center}}
\def\bfig{\begin{figure}}
\def\efig{\end{figure}}

\def\bit{\begin{itemize}}
\def\eit{\end{itemize}}

\def\f{\frac}

\def\[{\left[}
\def\]{\right]}
\def\({\left(}
\def\){\right)}

\def\..{\left.}
\def\.{\right.}
\def\tl{\tilde}
\def\ra{\rightarrow}
\def\la{\leftarrow}

\def\da{\dagger}

\def\la{\lambda}

\def\ep{\epsilon}

\def\De{\Delta}

\renewcommand{\thefootnote}{\fnsymbol{footnote}}

\title{Triplet fermion WIMP dark matter in the general supersymmetric Georgi-Machacek model}
\author[1]{Liangliang Shang,}
\author[1,2]{Yuanping Wang,}
\author[1,2]{Xiaokang Du\footnotemark[1]\footnotetext[1]{Corresponding author.},}
\author[1]{Bingfang Yang,}
\author[3,4]{Stefano Moretti}
\affiliation[1]{Centre for Theoretical Physics, Henan Normal University, Xinxiang 453007, P. R. China}
\affiliation[2]{Institute of Physics, Henan Academy of Sciences, Zhengzhou 450046, P. R. China}
\affiliation[3]{Department of Physics and Astronomy, Uppsala University, Box 516, SE-751 20 Uppsala, Sweden}
\affiliation[4]{School of Physics and Astronomy, University of Southampton, Highfield, Southampton SO17 1BJ, UK}
\emailAdd{xkdu@hnas.ac.cn, shangliangliang@htu.edu.cn, wangyuanping@hnas.ac.cn, yangbingfang@htu.edu.cn, s.moretti@soton.ac.uk; stefano.moretti@physics.uu.se}

\abstract{The supersymmetric custodial triplet model (SCTM), which is a fully-super\-symmetric generalization of the Georgi-Machacek (GM) model, is constructed by extending the Higgs sector of the minimal supersymmetric standard model by three triplet chiral superfields with hypercharge $Y=0,\pm 1$, in order to maintain the holomorphy of the superpotential and satisfy the requirements of anomaly cancellation. The global $SU(2)_L\otimes SU(2)_R$ symmetry has to be respected for the superpotential and soft supersymmetry breaking sector, where the former will only be broken by the Yukawa couplings and the latter one will be broken spontaneously  to the custodial $SU(2)_V$ symmetry after electroweak symmetry breaking similar as the GM model. The ensuing complicated spectrum not only gives rich collider phenomenology but also provides more available Weakly Interacting Massive Particle (WIMP) Dark Matter (DM) candidates. In this paper, we explore the viability of WIMP solutions for DM in the SCTM, by considering the increasingly stringent constraints from direct detection DM experiments. Our numerical simulations show that a significant portion of the SCTM parameter space remains viable despite  these constraints and will be fully tested in future experiments.}

\begin{document}
\maketitle
\newpage
\section{Introduction}
\label{sec-1}

\renewcommand{\thefootnote}{\textsuperscript{\arabic{footnote}}}

The discovery in 2012 of a 125 GeV Higgs boson~\cite{ATLAS:higgs,CMS:higgs} was a big leap for  particle physics. In fact, following the precise measurements of its properties, it  was soon clear that such a new state of nature was very compatible with the predictions of the 
Standard Model (SM). However, the leftover theoretical and experimental uncertainties mean that the specific process triggering electroweak symmetry breaking (EWSB) is still an open question, so searching for extra Higgs fields beyond the SM remains one of the most important tasks of the Large Hadron Collider (LHC). Additionally,  unresolved  experimental and theoretical issues, such as dark matter (DM), whose existence has been widely accepted through numerous evidences confirmed by astronomy, and the gauge hierarchy problem, which is essentially unwanted fine-tuning of parameters,  suggest that there should be new physics beyond the SM.

The Georgi-Machacek (GM) model ~\cite{GM,GM2,GM3,GM4,GM5,GM6,GM7,GM8,GM9,GM10,GM11,GM12,GM13,GM14,GM15,GM16,GM17, GM18,GM19,GM20,GM21,GM22,GM23,GM24,GM25,GM26,GM27,GM28,GM29,GM30,GM31,GM32,GM33,GM34,GM35,GM36,GM37,GM38,GM39,GM40,GM41,GM42,GM43,GM44,GM45,GM46,GM47,GM48,GM49,GM50,GM51,GM52,GM53,GM54,GM55,GM56,GM57,GM58,GM59,GM60,GM61,GM62,GM63,GM64,GM65,GM66,GM67,GM68,GM69,GM70,GM71,GM73,GM74} extends the SM by adding a complex $SU(2)_L$ triplet scalar $\chi$ (hypercharge 1) and a real $SU(2)_L$ triplet scalar $\xi$ (hypercharge 0). The GM model can nontrivially allow custodial symmetry preservation at tree level after EWSB, provided that the vacuum expectation values (VEVs) of these triplets align precisely. This alignment enables the triplet state VEV to reach tens of GeV, higher than in other triplet-extended SM scenarios. The GM model also predicts larger couplings of the SM-like Higgs boson to $W$ and $Z$ gauge bosons as well as the existence of nine more physical (pseudo)scalars, including a doubly charged one, which is the unique interesting feature of the GM.

Supersymmetry (SUSY) \cite{SUSY1,SUSY2,SUSY3,SUSY4} is one of the most fashionable extensions to the SM, postulating a symmetry linking fermions and bosons. It essentially doubles the particle spectrum, which can then elegantly solve the aforementioned fine-tuning problem. In the minimal supersymmetric standard model (MSSM), one should also double the doublet Higgs field to maintain the holomorphy of the superpotential and satisfy the requirements of anomaly cancellations. These two doublet Higgs fields exactly give mass to up- and down-type quarks separately. Besides, the lightest supersymmetric particle (LSP) is a natural DM candidate if we assume that $R$ parity is conserved and that the LSP is uncharged and colorless.

It is well known that the GM model faces a more serious hierarchy problem than the SM for its complex (pseudo)scalar fields. The MSSM also runs into trouble, though, because a 125 GeV Higgs mass requires too heavy top squark and/or a significant trilinear coupling $A_t$ (the so-called little hierarchy problem) plus the only available DM candidate suffers from increasingly stringent constraints from DM direct detection experiments. To remedy these problems, the supersymmetric custodial triplet model (SCTM)~\cite{SCTM1,SCTM2,SCTM3,SCTM4,SCTM5,SCTM6,SCTM7,SCTM8,SCTM11,SCTM9,SCTM10}, a general fully-supersymmetric extension of the GM model\footnote{Compared to the SCTM, the supersymmetric Georgi-Machacek (SGM) model \cite{SCTM8,SCTM11} is derived from the SCTM by taking a specific limit of the model parameters. In this limit, the SGM model exhibits the same (pseudo)scalar particle spectrum as the conventional Georgi-Machacek (GM) model at low energies, while retaining the supersymmetric properties of the SCTM. Here, we only discuss the fully supersymmetric realization of the GM model, i.e., the SCTM.}, is constructed by adding three triplet chiral superfields with hypercharge $Y=0,\pm1$ to the Higgs sector of the MSSM . The global $SU(2)_L\otimes SU(2)_R$ symmetry must be maintained for both the superpotential and soft SUSY breaking sector.  The former is broken only by Yukawa couplings while the latter spontaneously breaks to the custodial $SU(2)_V$ symmetry after EWSB, similar to the GM model. The emerging scenario not only offers rich collider phenomenology but also provides more viable Weakly Interacting Massive Particle (WIMP) DM candidates.

In fact, the SCTM has a more complex mass spectrum and richer particle content by adding not only triplets with hypercharge $Y=0,+1$ from the GM model compared to MSSM, but also an additional triplet with hypercharge $Y=-1$ for the same reason as doubling the doublet scalar in the MSSM. 
Thanks to the new $F$-term contributions to the Higgs mass at tree level from the introduced triplets, the SCTM successfully accommodates the observed 125 GeV Higgs mass without generating the aforementioned little hierarchy problem, thus making the theory more natural.

In the MSSM, WIMP-type DM  faces significant challenges from direct detection constraints. The lightest neutralino, the leading DM candidate, must have a very small interaction cross section with nucleons to evade direct detection by experiments like LZ~\cite{LZ}, PandaX-4T~\cite{PandaX-4T}, and XENONnT~\cite{XENONnT}. However, the Majorana fermion components of the additional triplet superfields in the SCTM allow for additional DM candidates, through the tripletino field. For example,  in a realistic SCTM~\cite{SCTM3}, the lightest neutralino, dominantly binolike but mixed with Higgsinos or tripletinos, serves as a viable DM candidate. Key mechanisms for achieving the observed relic density include resonant $s$-channel annihilations via triplet-like (pseudo)scalars and well-tempering through bino-tripletino coannihilations. Direct detection prospects are suppressed due to custodial symmetry (minimizing $Z$-mediated spin-dependent interactions) and small Higgsino components (reducing Higgs-mediated spin-independent cross sections), with most viable points lying below current bounds from direct detection.

In this work, we would like to investigate the situation of the SCTM in which the WIMP candidate, the LSP, is the lightest tripletinolike neutralino. We explore whether there is any parameter space that can fulfill the most stringent direct detection experimental constraints, when considering the effect of the suppression of the coupling between the neutralino and the $Z$ boson by the custodial symmetry. Furthermore, we are also committed to finding parameter space to provide the correct relic abundance of DM via $s$-channel funnel annihilation or coannihilation mechanisms for tripletinolike DM.

The rest of this paper is organized as follows. In Sec.~\ref{sec-2}, we briefly review the GM model and the SCTM. The spectrum of Higgs and neutralino states is given in Sec.~\ref{sec-3}. In Sec.~\ref{sec-4}, we present the experimental constraints applied, the specific numerical calculations performed and our numerical results. Finally, we show our conclusion in Sec.~\ref{sec-5}.

\section{Georgi-Machacek Model and its Supersymmetric Version}
\label{sec-2}

\subsection{A brief review of the Georgi-Machacek model}
\label{sec-21}
The Higgs sector of the GM model contains not only the SM-like $SU(2)_L$ complex doublet Higgs field $({\phi}^+,~{\phi}^0)$ carrying a $1/2$ hypercharge but also a $SU(2)_L$ complex triplet Higgs field $({\chi}^{++},~{\chi}^{+},~{\chi}^{0})$ with a unit hypercharge and a $SU(2)_L$ real triplet Higgs field $({\xi}^{+},~{\xi}^{0},~\xi^-)$ possessing a zero hypercharge. 
The most general scalar potential of the GM model being invariant under the global $SU(2)_L \times SU(2)_R$ symmetry and SM gauge symmetry can be given as
\begin{align}
V(\Phi,\De) & =  \frac{1}{2} m_{\Phi}^2 {\rm Tr} \left[ \Phi^\da \Phi \right] + \frac{1}{2} m_{\De}^2 {\rm Tr} \left[ \De^\da \De \right] + \la_1 \( {\rm Tr} \left[ \Phi^\da \Phi \right] \)^2 + \la_2 \( {\rm Tr} \left[ \De^\da \De \right] \)^2 \notag \\
& + \la_\De {\rm Tr} \left[ \( \De^\da \De \)^2 \right] + \la_4 {\rm Tr} \left[ \Phi^\da \Phi \right] {\rm Tr} \left[ \De^\da \De \right] + \la_5 {\rm Tr} \left[ \Phi^\da \frac{\sigma^a}{2} \Phi \frac{\sigma^b}{2} \right] {\rm Tr} \left[ \De^\da T^a \De T^b \right] \notag \\
& + \mu_1 {\rm Tr} \left[ \Phi^\da \frac{\sigma^a}{2} \Phi \frac{\sigma^b}{2} \right] (P^\da \De P)_{ab} + \mu_2 {\rm Tr} \left[ \De^\da T^a \De T^b \right] (P^\da \De P)_{ab},
\label{eq:GMpot}
\end{align}
where the $\Phi$ and $\De$ are the matrix form of doublet and triplet fields to tell whether the custodial symmetry is preserved or not more easily, and the specific representation can be shown as
\begin{align}
\Phi=\(
\begin{array}{cc}
\phi^{0*} & \phi^+ \\
\phi^- & \phi^0
\end{array}\),\quad
\De=\(
\begin{array}{ccc}
\chi^{0*} & \xi^+ & \chi^{++} \\
\chi^- & \xi^0 & \chi^{+} \\
\chi^{--} & \xi^- & \chi^{0}
\end{array}\)~.
\label{eq:Higgs_matrices}
\end{align}
The $\sigma^a~\(a=1,2,3\)$  (the Pauli matrices) in the potential are the generators of the $\(\textbf{2},\bold{\bar{2}}\)$ representation of the $SU(2)$ group while the $T^a~\(a=1,2,3\)$ are the generators for the $\(\textbf{3},\bold{\bar{3}}\)$ representation: 
\begin{align}
T^1 = & \frac{1}{\sqrt{2}} \( \begin{array}{ccc} 0 & 1 & 0 \\ 1 & 0 & 1 \\ 0 & 1 & 0 \end{array} \),~T^2=\frac{1}{\sqrt{2}} \( \begin{array}{ccc} 0 & -i & 0 \\ i & 0 & -i \\ 0 & i & 0 \end{array} \),~T^3 = \( \begin{array}{ccc} 1 & 0 & 0 \\ 0 & 0 & 0 \\ 0 & 0 & -1 \end{array} \).
\end{align}
Additionally,  $P$ is a matrix to rotate the triplet into its Cartesian basis
\begin{align}
P = & \frac{1}{\sqrt{2}} \( \begin{array}{ccc} -1 & i & 0 \\ 0 & 0 & \sqrt{2} \\ 1 & i & 0 \end{array} \).
\end{align}

The neutral real scalar field can be parametrized as ${\xi}^0  \ra v_{\xi}+h_{\xi}$ and the neutral components of complex scalar fields are decomposed into real and imaginary parts as
\beqa
{\phi}^0  \ra \(v_{\phi}+h_{\phi}+i a_{\phi}\)/\sqrt{2},~~~{\chi}^0  \ra v_{\chi}+\(h_{\chi}+i a_{\chi}\)/\sqrt{2}.
\eeqa 
The parameters $v_{\xi}$, $v_{\phi}$ and $v_{\chi}$ are VEVs of the corresponding neutral scalar fields that trigger the breaking of the $SU(2)_L \times SU(2)_R$ symmetry. Especially, the custodial $SU(2)_V$ symmetry would be preserved at tree level if the triplet VEVs are aligned as $v_{\chi}=v_{\xi}=v_{\De}$. In other words, the VEVs of the scalar fields take the identity form
\beqa
\langle\Phi\rangle = \frac{1}{\sqrt{2}}\( \begin{array}{cc} v_{\phi} & 0\\ 0 & v_\phi \end{array} \),~\langle\De\rangle=\( \begin{array}{ccc} v_\De & 0 & 0 \\ 0 & v_\De & 0 \\ 0 & 0 & v_\De \end{array} \).
\eeqa 
Henceforth, the EWSB condition becomes
\beqa
v^2=v_\phi^2+4 v_\chi^2+4 v_\xi^2=v_\phi^2+8 v_\De^2=\f{1}{\sqrt{2}G_F}\approx \(246~{\rm GeV}\)^2~.
\eeqa
Finally, the electroweak (EW) parameter $\rho$ in the GM model at tree level can be calculated as 
\beqa
\rho\equiv1+\De\rho ,~~~\De\rho=\frac{4(v_\xi^2-v_\chi^2)}{v^2}.
\eeqa
It is easy to tell that the scalar part of the GM model is indeed invariant under the custodial symmetry at tree level after EWSB. 
This alignment allows the triplet state VEV to reach tens of GeV, which is higher than in other triplet-extended SM models. Moreover, the GM model predicts enhanced couplings between the SM-like Higgs boson and the $W$ and $Z$ gauge bosons, along with the presence of nine additional physical (pseudo)scalars, including a doubly charged scalar particle—an intriguing and distinctive feature of the GM model, as previously mentioned.

\subsection{A brief review of the supersymmetric Georgi-Machacek model}
\label{sec-22}
The supersymmetric Georgi-Machacek model was proposed to address the  hierarchy problem of the GM model and embeds the useful custodial symmetry for the Higgs sector at tree level, in short, giving the SCTM.
The SCTM has the same Higgs doublet superfields as the MSSM with hypercharge $Y=(-1/2,~1/2)$,
\beqa
H_u=\(\begin{array}{c} H_u^+ \\ H_u^0\end{array}\), \quad H_d=\(\begin{array}{c} H_d^0 \\ H_d^-\end{array}\) 
\eeqa
and additional three  $SU(2)_L$ triplets with heypercharge $Y=(-1,~0,~1)$\footnote{Considering the holomorphic principle, which states that the superpotential can only depend on chiral superfields, not their complex conjugates \cite{Seiberg:1993vc}, the MSSM doubles the Higgs doublet scalar fields $H$ of the SM as $H_u,~H_d$ with opposite hypercharge for giving masses to up- and down-type fermions separately. For the same reason, the SCTM has to double the Higgs triplet scalar fields with nonzero hypercharge of the GM model, which results in three Higgs triplet superfields in the SCTM.}
\begin{equation}
\label{sigma}
\Sigma_{-}=\(\begin{array}{cc}
\frac{\chi^{-}}{\sqrt{2}} & \chi^0 \\
\chi^{--} & -\frac{\chi^{-}}{\sqrt{2}}
\end{array}\), \quad \Sigma_0=\(\begin{array}{cc}
\frac{\phi^0}{\sqrt{2}} & \phi^{+} \\
\phi^{-} & -\frac{\phi^0}{\sqrt{2}}
\end{array}\), \quad \Sigma_+=\(\begin{array}{cc}
\frac{\psi^{+}}{\sqrt{2}} & \psi^{++} \\
\psi^0 & -\frac{\psi^{+}}{\sqrt{2}}
\end{array}\).
\end{equation}

For the Higgs sector of SCTM to be invariant under the $SU(2)_R\times SU(2)_L$ symmetry, the Higgs scalar fields can be reorganized into the formation of bidoublet and bitriplet
\begin{equation}
\bar{H}=\(\begin{array}{c} H_d \\ H_u \end{array}\), 
\quad 
\bar{\De}=\(\begin{array}{cc} -\frac{\Sigma_0}{\sqrt{2}} & -\Sigma_{-} \\ -\Sigma_+ & \frac{\Sigma_0}{\sqrt{2}}
\end{array}\).
\label{bi-fields}
\end{equation}
The superpotential of the SCTM $W_{\rm SCTM}$ is made up by a Yukawa part $W_{MSSM/\mu}$ and a $SU(2)_R\times SU(2)_L$ invariant Higgs part $W_0$
\begin{equation}
\label{Higgs part superpotential}
W_0=\la \bar{H} \cdot \bar{\De} \bar{H}+\frac{\la_\De}{3} \rm{Tr} (\bar{\De}\bar{\De}\bar{\De})+\frac{\mu}{2} \bar{\it H} \cdot \bar{\it H}+\frac{\mu_{\De}}{2} \rm{Tr} (\bar{\De}\bar{\De}),
\end{equation}
where the antisymmetric dot products are defined by 
\beqa
\bar A\cdot \bar B\equiv \epsilon^{a b} \epsilon_{i j} \bar A_a^i \bar B_b^j, ~~~\epsilon^{12}=-\epsilon_{12}=1.
\eeqa

The total scalar potential is then
\beqa
V_{\rm SCTM}=V_F+V_D+V_{\rm soft},
\eeqa
where the $F$-term potential can be given as
\beqa
\begin{aligned}
V_F&= \left| \frac{\partial W}{\partial \Phi_a} \right|^2 \\ \notag
   &= \mu^2 \bar{H}^{\da} \bar{H} + \mu_{\De}^2 \rm{Tr} \[ \bar{\De}^{\da} \bar{\De} \] + 2 \la \mu \( \bar{\it H}^{\da} \bar{\De} \bar{\it H} + \text{ c.c. } \) \\
   &+ \la^2 \left\{ 4 \rm{Tr} \left[ \( \bar{\De} \bar{\it H} \)^{\da} \bar{\De} \bar{\it H} \right] + \( \bar{\it H}^{\da} \bar{\it H} \)^2 - \frac{1}{4} | \bar{\it H} \cdot \bar{\it H} |^2 \right\} \\
   &+ \la_{\De}^2 \left\{ \rm{Tr} \[ \bar{\De}^{\da} \bar{\De}^{\da} \bar{\De} \bar{\De} \] - \frac{1}{4} \rm{Tr} \[ \bar{\De}^{\da} \bar{\De}^{\da} \] \rm{Tr} \[ \bar{\De} \bar{\De} \] \right\} \\
   &+ \la \la_{\De} \left\{ \bar{H} \cdot \bar{\De}^{\da} \bar{\De}^{\da} \bar{H} - \frac{1}{4} \bar{H} \cdot \bar{H} \rm{Tr} \[ \bar{\De}^{\da} \bar{\De}^{\da} \] + h.c.  \right\} \\
   &+ \la \mu_{\De} \( \bar{H} \cdot \bar{\De}^{\da} \bar{H} + \text{ c.c. } \) + \la_{\De} \mu_{\De} \left\{ \rm{Tr} \[ \bar{\De}^{\da} \bar{\De}^{\da} \bar{\De} \] + h.c.  \right\}.
\end{aligned}
\eeqa
The $D$-term potential is
\beqa
V_D=-\(\frac{g_1}{2}+\frac{g_2}{2}\)\left\{\bar{H}^{\da} {\it Y}_H \bar{H}+\rm{Tr}\[\bar{\De} {\it Y}_{\De} \bar{\De}\]\right\}.
\eeqa
where  $g_1$ and $g_2$ are the gauge coupling constants associated with the $U(1)_Y$ and $SU(2)_L$ gauge groups, respectively.
The soft SUSY breaking potential is invariant under $SU(2)_R\times SU(2)_L$ and can be written  as
\beqa
\begin{aligned}
\label{soft potential}
V_{\rm soft} & =m_H^2\bar{H}^\da\bar{H}+m_{\De}^2 \rm{Tr}\[\bar{\De}^\da\bar{\De}\]+\frac{1}{2} {\it B}_{\mu} \bar{\it H} \cdot \bar{\it H} \\ 
& +\left\{\frac{1}{2} B_{\De} \rm{Tr} \[\bar{\De}\bar{\De}\]+{\it A}_{\la} \bar{\it H} \cdot \bar{\De} \bar{\it H}+\frac{1}{3} {\it A}_{\De} \rm{Tr}\[\bar{\De}\bar{\De}\bar{\De}\]+ H.c. \right\}.
\end{aligned}
\eeqa

After the neutral components of complex scalar fields develop VEVs to trigger the EWSB, the $SU(2)_L \times SU(2)_R$ symmetry will be broken into the custodial $SU(2)_V$ symmetry.
The neutral components of complex scalar fields are decomposed into real and imaginary parts as
\beqa
\Phi^0=\(v_{\Phi^0}+h_{\Phi^0}+i a_{\Phi^0}\)/\sqrt{2},~~~\Phi^0=H_u^0,~H_d^0,~\phi^0,~\chi^0,\psi^0.
\eeqa
For simplicity, we denote these VEVs as 
\beqa
v^2_{u}=v^2_{H_u^0},~~~v^2_{d}=v^2_{H_d^0},~~~v^2_{\phi}=v^2_{\phi^0},~~~v^2_{\chi}=v^2_{\chi^0},~~~v^2_{\psi}=v^2_{\psi^0}.
\eeqa 
Then the EW vacuum can be parametrized  as
\beqa
v^2\equiv v^2_u+v^2_d+v^2_{\phi}+v^2_{\chi}+v^2_{\psi}=(246 ~{\rm{GeV}})^2
\eeqa 
and the EW parameter $\rho$ in the SCTM at tree level can be given as 
\beqa
\De \rho = \frac{2(2 v^2_{\phi}-v^2_{\chi}-v^2_{\psi})}{v_u^2+v_d^2+2(2 v^2_{\phi}+v^2_{\chi}+v^2_{\psi})}.
\eeqa 
So, if we take the alignment limits $v^2_{\phi}=v^2_{\chi}=v^2_{\psi} \equiv v^2_\De$, the custodial symmetry will be preserved automatically for the Higgs sector at tree level\footnote{It is easy to see that the custodial symmetry will also be preserved at tree level when $2 v^2_{\phi}=v^2_{\chi}+v^2_{\psi}$, although this is not the typical GM model. We will investigate this scenario in our future work.}.

\section{Mass Spectrum of Higgs and Neutralino States}
\label{sec-3}

\subsection{Mass spectrum of Higgs states}
\label{sec-31}

The tadpole equations of the SCTM are
\beqa
\frac{\partial V}{\partial v_u}=\frac{\partial V}{\partial v_d}=\frac{\partial V}{\partial v_\phi}=\frac{\partial V}{\partial v_\psi}=\frac{\partial V}{\partial v_\chi}=0.
\eeqa
Then the minimization conditions can be derived as
\beqa
\begin{aligned}
M_{H_d}^2 &= \sqrt{2}v_\De(\la\mu-A_{\la}-\la\mu_\De)-\frac{v_\De^2}{2}(2\la_{\De}\la+5\la^2) + \tan{\beta}\left[ B_{\mu}+\frac{v_\De}{\sqrt{2}}(4\la\mu-A_{\la}-\la\mu_\De)\right. \\
&\left. -\frac{v_\De^2}{2}(\la_{\De}\la+4\la^2) \right] -\frac{g_1^2+g_2^2}{4}v_H^2\cos{2\beta}-v_H^2\la^2(\cos{\beta}^2+1)-\mu^2, \\
M_{H_u}^2 &= \sqrt{2}v_\De(\la\mu-A_{\la}-\la\mu_\De)-\frac{v_\De^2}{2}\(2\la_{\De}\la+5\la^2\) + \frac{1}{\tan{\beta}}\left[B_{\mu}+\frac{v_\De}{\sqrt{2}}(4\la\mu-A_{\la}-\la\mu_\De) \right. \\
& \left. -\frac{v_\De^2}{2}(\la_{\De}\la+4\la^2)\right] -\frac{g_1^2+g_2^2}{4}v_H^2\cos{2\beta}-v_H^2\la^2(\sin{\beta}^2+1)-\mu^2, \\
%
%
M_{\Sigma_+}^2&=-\left\{B_\De+\mu_\De^2+\frac{v_\De}{\sqrt{2}}\left(A_\De+3\la_\De\mu_\De\right)+\frac{1}{\sqrt{2}}\frac{v_H^2}{v_\De}\left(2A_\la\cos^2{\beta}+2\la\mu_\De \sin^2{\beta}-4\cos{\beta} \sin{\beta}\la\mu\right) \right. \\
& \left. +\frac{v_H^2}{2}\[-(g_1^2+g_2^2)\cos{2\beta}+2\cos{\beta} \sin{\beta}(\la_\De\la+2\la^2)+2\sin^2{\beta}\la_\De\la+8\cos^2{\beta}\la^2\]+v_\De^2\la_\De^2\right\}, \\
M_{\Sigma_0}^2&=-\left\{B_\De+\mu_\De^2+\frac{v_\De}{\sqrt{2}}\left(A_\De+3\la_\De\mu_\De\right) \right.\\
&\left. +\frac{1}{\sqrt{2}}\frac{v_H^2}{v_\De}\[ 2\cos{\beta} \sin{\beta}(A_\la+\la\mu_\De)-2\la\mu \]+\frac{v_H^2}{2}\[8 \cos{\beta} \sin{\beta} \la^2+(2\la_\De\la+2\la^2)\]+v_\De^2\la_\De^2\right\}, \\
M_{\Sigma_-}^2&=-\left\{B_\De+\mu_\De^2+\frac{v_\De}{\sqrt{2}}\left(A_\De+3\la_\De\mu_\De\right)+\frac{1}{\sqrt{2}}\frac{v_H^2}{v_\De}\left(2A_\la\sin^2{\beta}+2\la\mu_\De \cos^2{\beta}-4\cos{\beta} \sin{\beta}\la\mu\right) \right.\\
&\left. +\frac{v_H^2}{2}\[-(g_1^2+g_2^2)\cos{2\beta}+2\cos{\beta} \sin{\beta}(\la_\De\la+2\la^2)+2\cos^2{\beta}\la_\De\la+8\sin^2{\beta}\la^2\]+v_\De^2\la_\De^2\right\}, 
\label{minimization conditions }
\end{aligned}
\eeqa
where the mixing angle of the Higgs doublets is defined as 
\beqa
\tan{\beta}=\frac{v_u}{v_d},
\eeqa
and we have $v_H=\sqrt{(v_u^2+v_d^2)/2}.$

The bidoublets and the bitriplets~(\ref{bi-fields}) can be decomposed as $(\mathbf{2}, \mathbf{2})=\mathbf{1} \oplus \mathbf{3}$ and $(\mathbf{3}, \mathbf{3})=\mathbf{1} \oplus \mathbf{3} \oplus \mathbf{5}$ to provide the eigenstates of the $SU(2)_V$ custodial symmetry for the  Higgs fields after EWSB, which can then  be rotated to the mass eigenstates with mixing matrices.
\begin{itemize}
    \item  {$SU(2)_V$ Singlets} \\
    There are two neutral scalar singlets 
    \begin{equation}
    \begin{aligned}
    h_1=\frac{1}{\sqrt{2}}\(H_d^0+H_u^0\),~~~\delta_1 =\frac{\phi^0+\chi^0+\psi^0}{\sqrt{3}},
    \end{aligned}
    \end{equation}
    which can be rotated to mass eigenstates of two $CP$-even singlets $(s_1,~s_2)$ and two $CP$-odd singlets $(a_1,~a_2)$ 
    \begin{equation}
    \begin{aligned}
        \(s_{1}, s_{2}\) \(\begin{array}{cc} M_{s_1}^2 & 0 \\0 & M_{s_2}^2 \end{array}\)  \(\begin{array}{c} s_{1} \\ s_{2} \end{array}\) =
        \(h_{1}^{r}, \delta_{1}^{r}\) \mathcal{M}_S^2 \(\begin{array}{c} h_{1}^{r} \\ \delta_{1}^{r} \end{array}\), \\
        \(a_{1}, a_{2}\) \(\begin{array}{cc} M_{a_1}^2 & 0 \\0 & M_{a_2}^2 \end{array}\)  \(\begin{array}{c} a_{1} \\ a_{2} \end{array}\) =
        \(h_{1}^{i}, \delta_{1}^{i}\) \mathcal{M}_A^2 \(\begin{array}{c} h_{1}^{i} \\ \delta_{1}^{i} \end{array}\).
    \end{aligned}
    \end{equation}
    The matrix elements are \begin{equation}
    \begin{aligned}
    & (\mathcal{M}_S)^2_{11}=6 \la^2 v_H^2, \\
    & (\mathcal{M}_S)^2_{22}=\frac{v_H^2\left[\la\(2 \mu-\mu_{\De}\)-A_{\la}\right]+v_{\De}^2\left[-A_\De+\la_\De\(4 \la_\De v_{\De}-3 \mu_{\De}\)\right]}{v_{\De}}, \\
    & (\mathcal{M}_S)^2_{12}=\(\mathcal{M}_S^2\)_{21}=\sqrt{6} v_H\left[A_{\la}+\la\(6 \la v_{\De}-2 \la_\De v_{\De}-2 \mu+\mu_{\De}\)\right];\\
    & (\mathcal{M}_A)^2_{11}=2\(B_{\mu}-3 v_{\De}\left[A_{\la}+\la\(-\la_\De v_{\De}+\mu_{\De}\)\right]\), \\
    & (\mathcal{M}_A)^2_{22}=-\frac{v_H^2\(A_{\la}-2 \la \mu\)+v_{\De}\(-3 A_\De v_{\De}+2 B_{\De}-4 \la \la_\De v_H^2\)+\(\la v_H^2-\la_\De v_{\De}^2\) \mu_{\De}}{v_{\De}}, \\
    & (\mathcal{M}_A)^2_{12}=\(\mathcal{M}_A^2\)_{21}=\sqrt{6} v_H\left[\la\(-2 \la_\De v_{\De}+\mu_{\De}\)-A_{\la}\right].
    \end{aligned}
    \end{equation}

    \item  {$SU(2)_V$ Triplets} \\ 
    There are two $CP$-even scalar triplets and also two $CP$-odd scalar triplets
    \begin{equation}
    \begin{aligned}
    &h_3^{+}=H_u^{+}, \quad h_3^0=\frac{1}{\sqrt{2}}\(H_d^0-H_u^0\), \quad h_3^{-}=H_d^{-},\\
    \delta_3^{+} & =\frac{\psi^{+}-\phi^{+}}{\sqrt{2}}, \delta_3^0=\frac{\chi^0-\psi^0}{\sqrt{2}}, \delta_3^{-}=\frac{\phi^{-}-\chi^{-}}{\sqrt{2}}.
    \end{aligned}
    \end{equation}
    The $CP$-even scalar triplets
    \begin{equation}
    T_H=\(\begin{array}{c} \frac{1}{\sqrt{2}}\(h_3^{+}+h_3^{-*}\) \\ h_{3}^{0,r} \\ \frac{1}{\sqrt{2}}\(h_3^{-}+h_3^{+*}\) \end{array}\), \quad 
    T_{\De}=\(\begin{array}{c} \frac{1}{\sqrt{2}}\(\delta_3^{+}+\delta_3^{-*}\) \\ \delta_{3}^{0,r} \\ \frac{1}{\sqrt{2}}\(\delta_3^{-}+\delta_3^{+*}\) \end{array}\),
    \end{equation}
    are mixed by a squared mass matrix $\mathcal{M_T}^2$,
    \begin{equation}
    \(T_1, T_2\)\(\begin{array}{cc} M_{T_1}^2 & 0 \\ 0 & M_{T_2}^2
    \end{array}\)\(\begin{array}{c} T_1 \\ T_2 \end{array}\)
    =
    \(T_H, T_\De\)\(\begin{array}{cc} \mathcal{M}_{T_{11}}^2 &\mathcal{M}_{T_{12}}^2 \\ \mathcal{M}_{T_{21}}^2 & \mathcal{M}_{T_{22}}^2 \end{array}\) \(\begin{array}{c} T_H \\ T_\De \end{array}\),
    \end{equation}
    The matrix elements are 
    \begin{equation}
    \begin{aligned}
    \mathcal{M}_{T_{11}}^2 & =g^2 v_H^2+2 \la\(-4 \la v_{\De}^2+\la v_H^2+4 \mu v_\De\)+2 B_{\mu}-2 v_\De\left[A_{\la}+\la\(\mu_{\De}-\la_\De v_{\De}\)\right], \\
    \mathcal{M}_{T_{22}}^2 & =4 g^2 v_\De-\left[2 B_{\De}-3 \la \la_\De v_H^2+2 v_\De\(\la_\De^3 v_{\De}-A_\De\)\right] \\
    & -\frac{1}{v_\De}\left[2 \la v_H^2\(\la v_\De-\mu\)-\(\la v_H^2-2 \la_\De v_\De^2\) \mu_{\De}-v_H^2 A_{\la}\right] \\
    \mathcal{M}_{T_{12}}^2 & =\mathcal{M}_{T_{21}}^2=2 v_H\left[-A_{\la}+v_\De\(g^2-4 \la^2-\la \la_\De\)+\la \mu_{\De}\right],
    \end{aligned}
    \end{equation}
    where $g=g_2$ for charged fields and $g=\sqrt{g_1^2+g_2^2}$ for neutral fields.

    For two $CP$-odd pseudoscalar triplets,
    \begin{equation}
    \begin{aligned}
    & G_3^0 =\cos{\alpha_T} h_{3}^{0,i}+\sin{\alpha_T} \delta_{3}^{0,i},~~~~~G_3^{\mp} =\cos{\alpha_T}\frac{h_3^{ \pm *}-h_3^{\mp}}{\sqrt{2}}+\sin{\alpha_T} \frac{\delta_3^{ \pm *}-\delta_3^{\mp}}{\sqrt{2}},\\
    & A_3^0 =-\sin{\alpha_T} h_{3}^{0,i}+\cos{\alpha_T} \delta_{3}^{0,i},~~~A_3^{\mp} =-\sin{\alpha_T} \frac{h_3^{ \pm *}-h_3^{\mp}}{\sqrt{2}}+\cos{\alpha_T} \frac{\delta_3^{ \pm *}-\delta_3^{\mp}}{\sqrt{2}},
    \end{aligned}
    \end{equation}
    where the parameter $\alpha_T$ is defined as the mixing angle of custodial triplets
    \begin{equation}
    \sin{\alpha_T}=\frac{2 \sqrt{2} v_{\De}}{v}, \quad \cos{\alpha_T}=\frac{\sqrt{2} v_H}{v} .
    \end{equation}
    The pseudoscalars $G^{0,\pm}_3$ are the massless Goldstone bosons absorbed by $W$ and $Z$. The mass of the pseudoscalar $A_3$ can be written as
    \begin{equation}
    m_A^2=\frac{v_H^2+4 v_{\De}^2}{v_{\De}}\(\la\left[2 \mu-\mu_{\De}-\(2 \la-\la_\De\) v_{\De}\right]-A_{\la}\).
    \end{equation}

    \item  {$SU(2)_V$ Quintuplet} \\
    The quintuplets are composed of ditriplets only
    \begin{equation}
    \begin{aligned}
    \delta_5^{++} & =\psi^{++}, \delta_5^{+}=\frac{\phi^{+}+\psi^{+}}{\sqrt{2}}, \delta_5^0=\frac{-2 \phi^0+\psi^0+\chi^0}{\sqrt{6}}, \delta_5^{-}=\frac{\phi^{-}+\chi^{-}}{\sqrt{2}}, \delta_5^{--}=\chi^{--},
    \end{aligned}
    \end{equation}
    being decomposed as one $CP$-even scalar quintuplet $Q_s$ and its mirror $CP$-odd pseudoscalar $Q_a$,
    \begin{equation}
    Q_s=\(\begin{array}{c}
    \frac{1}{\sqrt{2}}\(\delta_5^{++}+\delta_5^{--*}\) \\
    \frac{1}{\sqrt{2}}\(\delta_5^{+}+\delta_5^{-*}\) \\
    \delta_{5}^{0,r} \\
    \frac{1}{\sqrt{2}}\(\delta_5^{-}+\delta_5^{+*}\) \\
    \frac{1}{\sqrt{2}}\(\delta_5^{--}+\delta_5^{++*}\)
    \end{array}\), 
    \quad Q_a=\(\begin{array}{c}
    \frac{1}{\sqrt{2}}\(\delta_5^{++}-\delta_5^{--*}\) \\
    \frac{1}{\sqrt{2}}\(\delta_5^{+}-\delta_5^{-*}\) \\
    \delta_{5}^{0,i} \\
    \frac{1}{\sqrt{2}}\(\delta_5^{-}-\delta_5^{+*}\) \\
    \frac{1}{\sqrt{2}}\(\delta_5^{--}-\delta_5^{++*}\)
    \end{array}\).
    \end{equation}
    The masses are \begin{equation}
    \begin{aligned}
    & M_{Q_s}^2=\frac{v_H^2\left[\la\(2 \mu-\mu_{\De}\)-A_{\la}\right]}{\sqrt{2} v_{\De}}+\frac{3}{2} \la v_H^2\(\la_{\De}-2 \la\) \sqrt{2} v_{\De}\(3 \la_{\De} \mu_{\De}+A_{\De}\)-v_{\De}^2 \la_{\De}^2, \\
    & M_{Q_a}^2=\frac{v_H^2\left[\la\(2 \mu+\mu_{\De}\)+A_{\la}\right]}{\sqrt{2} v_0}-\frac{1}{2} v_H^2\(6 \la+\la_\De\)+2 \sqrt{2} \la_\De v_\De \mu_{\De}-2 B_{\De} .
    \end{aligned}
    \end{equation}
 
\end{itemize}

In the light of the above,  the SCTM has  five $CP$-even Higgs states $H_{1,2,3,4,5}$, four $CP$-odd Higgs states $A_{1,2,3,4}$, five charged Higgs states $H^\pm_{1,2,3,4,5}$, and two  doubly charged Higgs states $H^{\pm\pm}_{1,2}$.

\subsection{Mass spectrum of Neutralino states}
\label{sec-32}

The lightest neutralino, as the LSP, is an ideal WIMP DM candidate. Neutralinos of SCTM are formed by two gauginos (bino $\tilde{B}$ and wino $\tilde{W}_3$) and two Higgsinos ($\tilde{H}_u^0$ and $\tilde{H}_d^0$) and three more tripletinos ($\tilde{\phi}^0$, $\tilde{\chi}^0$ and $\tilde{\psi}^0$). In this basis
\beqa
\psi_N = \( \tilde{B}, \tilde{W}_{3} , \tilde{H}_d^0 , \tilde{H}_u^0 , \tilde{\phi}^0 , \tilde{\chi}^0 , \tilde{\psi}^0 \),
\eeqa
the mass matrix of neutralinos $\mathcal{M}_N$ can be given by

\small
\beqa
\(  \begin{array}{ccccccc}
M_1 & 0 & -\frac{1}{\sqrt{2}} g_1  \cos_{\beta} v_H & \frac{1}{\sqrt{2}} g_1 \sin_{\beta} v_H & 0 & - g_1 v_{\De} & g_1 v_{\De} \\
0 & M_2 & \frac{1}{\sqrt{2}} g_2 \cos_{\beta} v_H & -\frac{1}{\sqrt{2}} g_2  \sin_{\beta} v_H & 0 & g_2 v_{\De} & - g_2 v_{\De} \\
-\frac{1}{\sqrt{2}} g_1  \cos_{\beta} v_H & \frac{1}{\sqrt{2}} g_2  \cos_{\beta} v_H & - \sqrt{2} \la v_{\De} & m_{\tilde{H}} & - \la \sin_{\beta} v_H & 0 & -2 \la \cos_{\beta} v_H \\
\frac{1}{\sqrt{2}} g_1 \sin_{\beta} v_H & -\frac{1}{\sqrt{2}} g_2  \sin_{\beta} v_H & m_{\tilde{H}} & -\sqrt{2} \la v_{\De} & - \la  \cos_{\beta} v_H & -2 \la \sin_{\beta} v_H & 0 \\
0 & 0 & - \la \sin_{\beta} v_H &  -\la \cos_{\beta} v_H & \mu_{\De} & -\frac{1}{\sqrt{2}} \la_\De v_{\De} & -\frac{1}{\sqrt{2}} \la_\De v_{\De} \\
-g_1 v_{\De} & g_2 v_{\De} & 0 & -2 \la  \sin_{\beta} v_H & -\frac{1}{\sqrt{2}} \la_\De v_{\De} & 0 & m_{\tilde{\De}} \\
g_1 v_{\De} & -g_2 v_{\De} & -2\la  \cos_{\beta} v_H & 0 & -\frac{1}{\sqrt{2}} \la_\De v_{\De} & m_{\tilde{\De}} & 0
\end{array} \),
\eeqa
\normalsize  

where
\begin{align} 
m_{\tilde{H}} = - \frac{1}{\sqrt{2}} v_{\De} \la  + \mu , ~~~
m_{\tilde{\De}} = - \frac{1}{\sqrt{2}} v_{\De} \la_{\De}  + \mu_{\De},
\end{align} 
in addition,  $M_1$ and $M_2$ are the masses of the gauginos $\tilde{B}$ and $\tilde{W}_3$, respectively.

It is quite difficult to obtain the mass eigenstates of neutralinos because their mass matrix in the SCTM is nontrivial. The same multiplets under the $SU(2)_V$ custodial symmetry can be rotated into the custodial basis. Since $\tilde{B}$ is a custodial singlet and  $\tilde{W}_3$ is a custodial triplet, they can mix separately with the custodial singlet Higgsinos $(\tilde{h}_1,~\tilde{\delta}_1$) and custodial triplet Higgsinos ($\tilde{h}_3^0, ~\tilde{\delta}_3^0$), respectively. Then the content of the neutralino will be changed into three (approximately) custodial singlets $(\tilde{h}_1,\tilde{\delta}_1,\tilde{B})$, three triplets $(\tilde{W}_3, \tilde{h}_3^0, \tilde{\delta}_3^0)$, and one fiveplet ($\tilde{\delta}^0_5$) if we consider the neutralino mass matrix in the custodial basis. Then the basis of neutralinos will be changed into
\beqa
\Psi_N = \(\tilde{h}_1, \tilde{\delta}_1, \tilde{B}, \tilde{W}_3, \tilde{h}_3^0, \tilde{\delta}_3^0, \tilde{\delta}^0_5 \),
\eeqa
the corresponding mass matrix being

\small
\beqa
\label{eq:neutralino}
\(  \begin{array}{ccccccc}
\frac{3}{\sqrt{2}} \la v_\De - \mu & \sqrt{3} \la v_H & 0 & 0 & 0 & 0 & 0 \\
\sqrt{3} \la v_H & -\sqrt{2}\la_\De v_\De + \mu_\De & 0 & 0 & 0 & 0 & 0 \\
0 & 0 &  \frac{M_V^g}{g^2} & \frac{g_1 g_2 M_V}{g^2}  & 0 & 0 & 0 \\
0 & 0 &  \frac{g_1 g_2 M_V}{g^2}  &  \frac{M_V^g}{g^2}  &\sqrt{2}gv_H & \sqrt{2}gv_\De & 0 \\
0 & 0 & 0 & \sqrt{2}gv_H  & \frac{1}{\sqrt{2}} \la v_\De + \mu & -\sqrt{2}\la v_H & 0 \\
0 & 0 & 0 & \sqrt{2}gv_\De & -\sqrt{2}\la v_H & \frac{1}{\sqrt{2}} \la_\De v_\De - \mu_\De & 0 \\
0 & 0 & 0 & 0 & 0 &0 &  \frac{1}{\sqrt{2}} \la_\De v_\De + \mu_\De 
\end{array} \),
\eeqa

\normalsize  
where the parameters $M_V,~M_V^g$ and $g$ are defined as 
\beqa
M_V=M_2-M_1,~~~M_V^g=g_1^2 M_1 + g_2^2 M_2,~~~g=\sqrt{g_2^2 + g_1^2}.
\eeqa

Considering that pure Higgsino-like and pure winolike DM are very severely restricted by DM direct detection,  we should take large values for $\mu$ and $M_2$ to prevent them from being the dominant components of the light neutralino. Additionally, the binolike DM indeed can escape the direct detection constraints but will spoil the meaning of triplets introduced by SCTM since this situation can be achieved in the simplest supersymmetric model, i.e., the MSSM. To realize tripletinolike DM, we need look only at parameter space for which $\mu_\De<\mu,M_1,M_2$. These tripletinolike neutrinos may be degenerate in the case of $v_\De \ll v_H$.

\section{Numerical Results}
\label{sec-4}

\subsection{Scan strategy and constraints}
\label{sec-41}

This work focuses on investigating the phenomenology of tripletinolike DM within the SCTM, which is free from heavy squark behavior. To achieve this, we set the soft breaking mass for squarks (except the top squark) to be larger than 2 TeV. Thanks to the $F$-term contribution provided by triplets, the SM-like 125 GeV Higgs boson can rely less on heavy top quarks and their large trilinear coupling. 
Considering that the EW parameter $\rho$ remains unity at tree level even when the VEVs of the Higgs doublets are misaligned, we treat $\tan{\beta}$ as a free parameter in this work, rather than adopting the fixed values used in \cite{SCTM3}. Moreover, in light of the significantly more stringent collider bounds compared to those in \cite{SCTM3}, we relax the soft trilinear and bilinear couplings as well as the mass terms of the colored supersymmetric particles. This relaxation also enables a more thorough exploration of the ensuing phenomenology.
This allows the following parameter ranges to evade constraints from collider experiments
\beqa
M_3>2200~{\rm{GeV}},~~~~1500{{\rm{GeV}}}<M_{\tilde{\it Q}_3, \tilde{\it U}_3}<5000 ~{{\rm{GeV}}},~~~|A_{t,b,\tau}| < 5000 ~{\rm{GeV}}.
\eeqa
Additionally, the soft breaking masses for sleptons are chosen as follows:
\beqa
200~{\rm{GeV}}<M_{\tilde{L}_{3}}<M_{\tilde{L}_{1,2}}< 1000 ~{\rm{GeV}}.
\eeqa
The gaugino masses $M_1$ and $M_2$, mass dimension parameters  $\mu$ and $\mu_\De$ as well as the corresponding bilinear parameters are selected to satisfy
\beqa
M_2>M_1,~\mu>\mu_\De>100~{\rm{GeV}},~~~B_\mu,~B_\De>(50~{\rm{GeV}})^2,
\eeqa
so as to ensure the realization of tripletinolike DM.
Finally, the remaining free parameters $\la,~\la_\Delta$ and $\tan{\beta}$ are chosen as
\beqa
|\la|,~|\la_\Delta|<\sqrt{4\pi},~~~~2<\tan{\beta}<60.
\eeqa

We use the package \text{SPheno}4.0.5 \cite{SPheno1,SPheno2}, with the model file for the  SCTM\footnote{Actually, its actual implementation in SARAH has been provided by Keping Xie~\cite{SCTM8,SCTM9}.} generated by SARAH4.15.2 \cite{SARAH1,SARAH2,SARAH3,SARAH4,SARAH5}, to calculate the realistic particle spectrum, branching ratios (BRs), total widths, and other relevant quantities, as well as to incorporate  some experimental constraints. Given the rich phenomenology associated with so many superparticles and additional Higgs bosons in the SCTM, we not only utilize the HiggsBounds5.10.0 package \cite{HiggsBounds1,HiggsBounds2,HiggsBounds3,HiggsBounds4,HiggsTools} to ensure that our parameter choices comply with the bounds on additional Higgs bosons at colliders, but also apply the HiggsSignals2.6.2 package \cite{HiggsTools,HiggsSignals1,HiggsSignals2,HiggsSignals3} to verify that the lightest $CP$-even Higgs boson mimics the behavior of the SM-like Higgs boson. 

For the first goal of investigating the tripletinolike DM candidate,  we also calculate the DM relic density, in both its spin-dependent and spin-independent cross sections against nucleons, with the package micrOMEGAs \cite{micrOMEGAs1,micrOMEGAs2,micrOMEGAs3,micrOMEGAs4}, after  checking that the LSP is the tripletinolike neutralino.

During the numerical scan, we impose the following constraints.
\bit
\item   To guarantee the appearance of a minimum for the scalar potential, we ask the  determinant of the Hessian at the origin be negative. This condition for the leading order in small $v_{\Delta}$ in the custodial case can be expressed as ~\cite{SCTM3}
\be
\la(2\mu-\mu_{\Delta})-A_{\la} > 0~~\text{ and }~~ \frac{3}{2} v_H^2 \la^2 -2m_3^2 <0.
\label{eqn:custmincond}
\ee

\item   The lower bounds on the mass of sparticles are \cite{ATLAS:2022ihe, CMS:2020cur, ATLAS:2021yij, ATLAS:2020dsf, ATLAS:2020aci, ATLAS:2024fub, ATLAS:2019lff} \\
    stau mass: $m_{\tl{\tau}} \gtrsim 0.48$ TeV;\\
    gluino mass: $m_{\tl{g}} \gtrsim 2.4$ TeV;\\
    light top squark and bottom squark mass: $m_{\tl{t}_1} \gtrsim 1.25$, $m_{\tl{b}_1} \gtrsim 1.25$ TeV;\\
    degenerate masses of the first two generations of squarks: $m_{\tl{q}} \gtrsim 1.0 - 1.4$ TeV. 

\item   The light $CP$-even Higgs state should be the SM-like Higgs boson with mass around 125 GeV. In addition, the component of this SM-like Higgs boson has to be dominated by the $H_u$ and $H_d$, while the contribution of the triplet scalar also should be sizable.
  \beqa
  122 {\rm ~GeV}<M_h <128 {\rm ~GeV},~~~|ZH_{1,H_u}|^2+|ZH_{1,H_d}|^2>0.9,
   \eeqa
where $ZH_{1,i}$ represents the contributions of different components to the light $CP$-even Higgs boson.

\item The DM relic density should satisfy  Planck data \cite{Planck}:
    \beqa
    \Omega_{DM} h^2 = 0.1199\pm 0.0027.
    \eeqa
As the mass matrix \ref{eq:neutralino} shows, the masses of tripletinolike neutralinos are degenerate so that the contribution of coannihilation among the tripletinolike neutralinos can suppress the relic density of tripletinolike DM. Thus we set an upper limit for the relic density in our scan based on the above equation.

\item The masses of sleptons and electroweakinos (ewinos) are usually not so large, they have to satisfy the lower bounds from the large electron positron (LEP) unquestionably~\cite{LEP}
    \beqa
    m_{\tl{\chi}^\pm}>103.5 ~{\rm GeV}. 
    \eeqa
Although the lower mass bound of ewinos have reached hundreds GeV from LHC, these constraints are not suitable for our model, because the masses of tripletinolike neutralinos and charginos are too degenerate to be identified. We also consider the mass bound of sleptons from LHC. We apply the package SModelS2.3~\cite{MahdiAltakach:2023bdn, Kraml:2013mwa} to determine whether a sample is excluded or not by decomposing the model spectrum and converting it into simplified model topologies so as to compare it with results from the LHC. We consider these
typical processes $pp \to \tilde{\chi}_{1,2}^0\tilde{\chi}_1^\pm,\tilde{\chi}_1^+\tilde{\chi}_1^-,\tilde{l}_1\tilde{l}_1^*$ ($\tilde{l}_1$ stands for the lightest slepton) and use the package MadGraph\_aMC@NLO~\cite{Alwall:2011uj} to generate the cross sections of these processes as inputs for SModelS. (The UFO model file of the SCTM for MadGraph\_aMC@NLO is generated by SARAH.) 

\eit

Finally, due to the constraints from the oblique parameters $S,~T$, and $U$ as well as those on the couplings of triplets to quarks, the BRs of $B$-meson rare decays are naturally consistent with experimental values.

\subsection{Viable parameter space}
\label{sec-42}

\begin{figure}
\centering
\includegraphics[width=.32\textwidth]{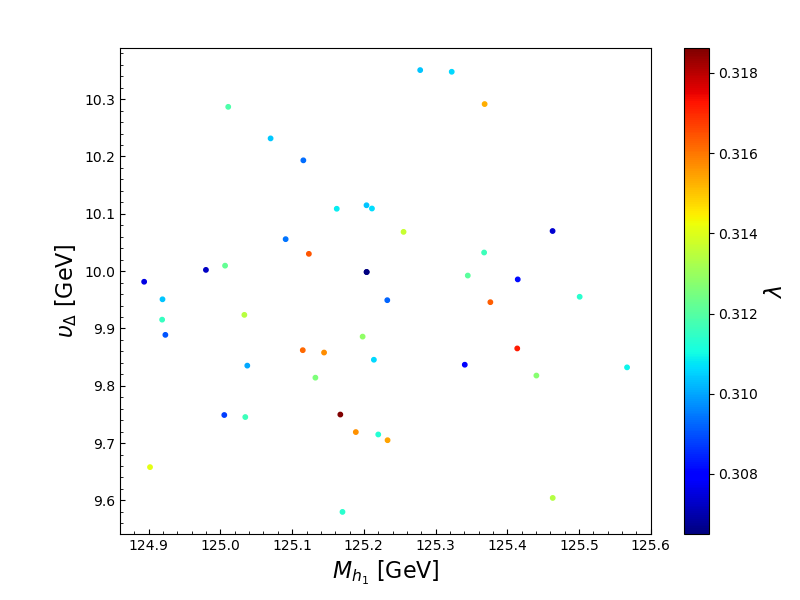}
\includegraphics[width=.32\textwidth]{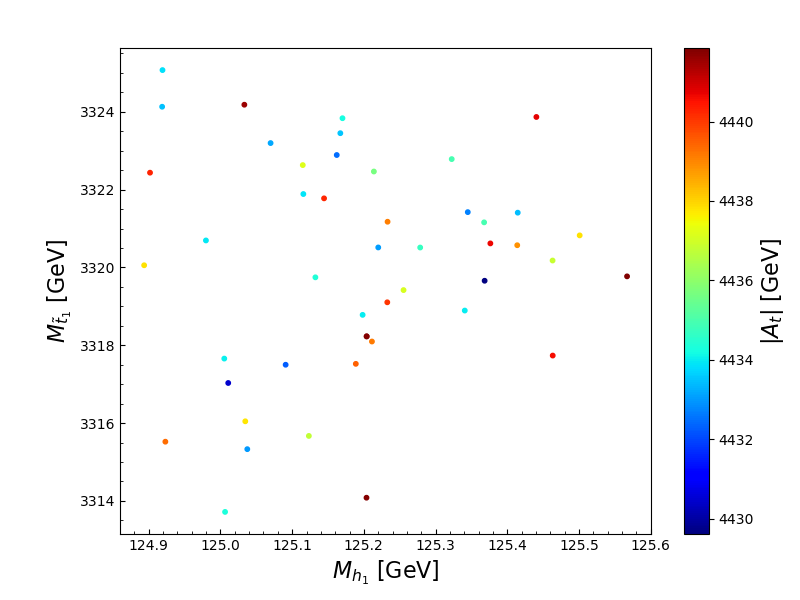}
\includegraphics[width=.32\textwidth]{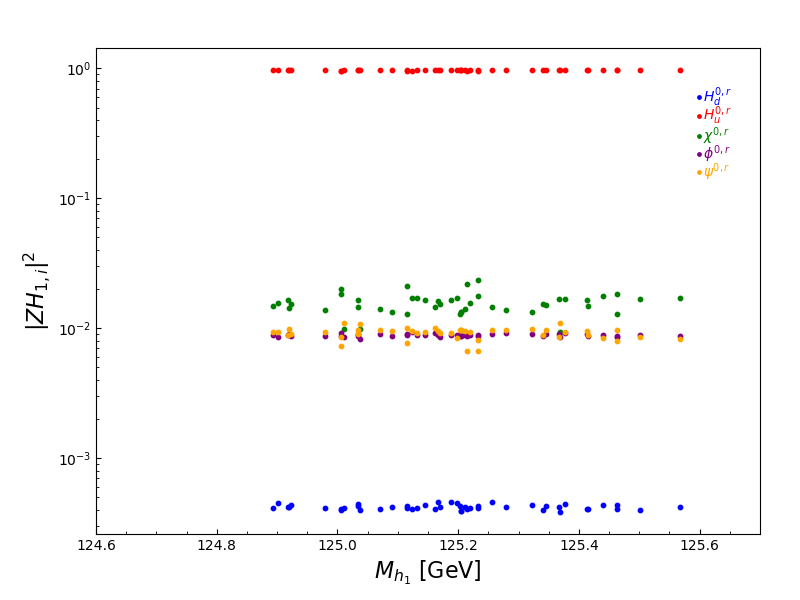}\\
\includegraphics[width=.32\textwidth]{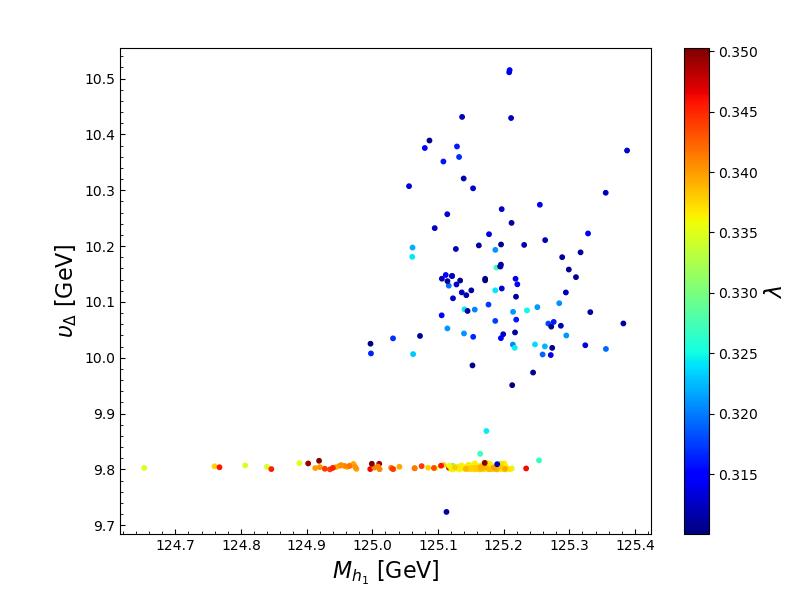}
\includegraphics[width=.32\textwidth]{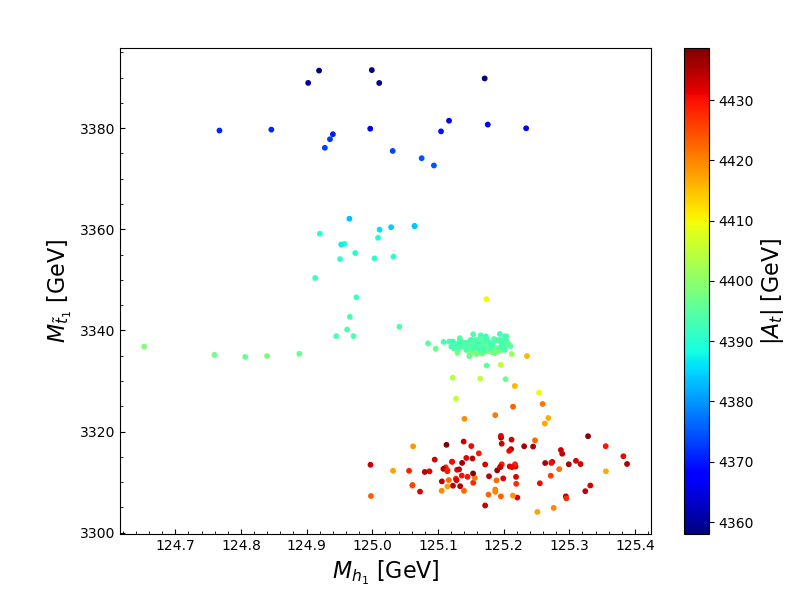}
\includegraphics[width=.32\textwidth]{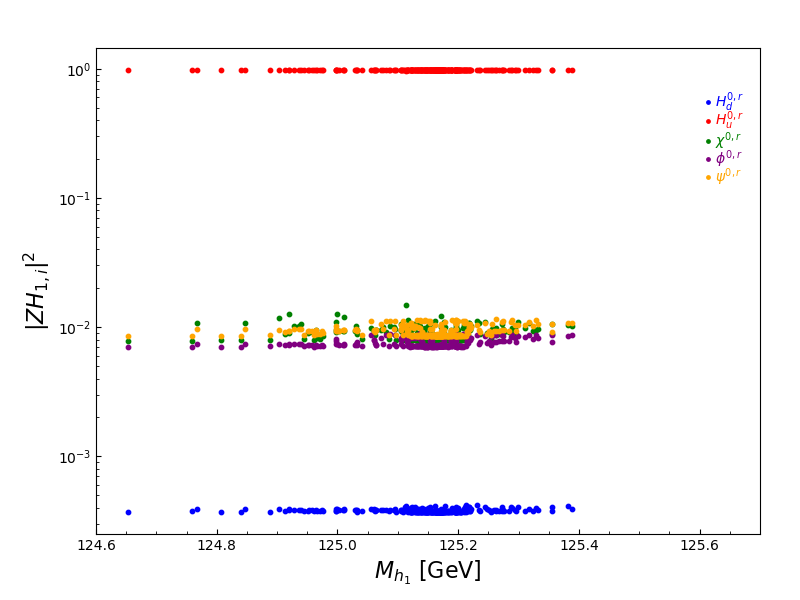}
\caption{Related information about Higgs boson of survived parameters. Left column: shows the relationship between the VEVs of triplets and the mass of the SM-like Higgs boson. The different colors denote the value of coupling parameter $\la$. The mass of top squark and the trilinear coupling $A_t$ can be found in the middle column. Right column: specific values of each component in the SM-like Higgs boson. 
}
\label{fig1}
\end{figure}

\begin{figure}
\centering
\includegraphics[width=.32\textwidth]{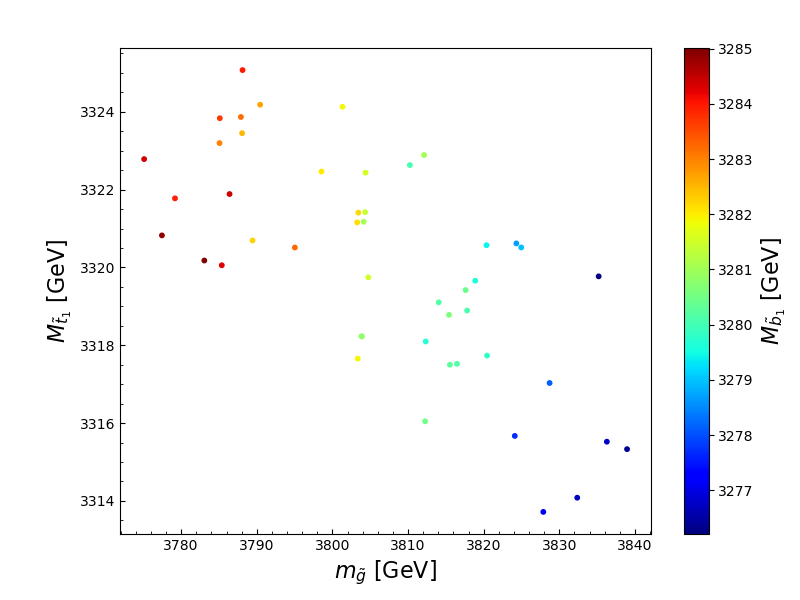}
\includegraphics[width=.32\textwidth]{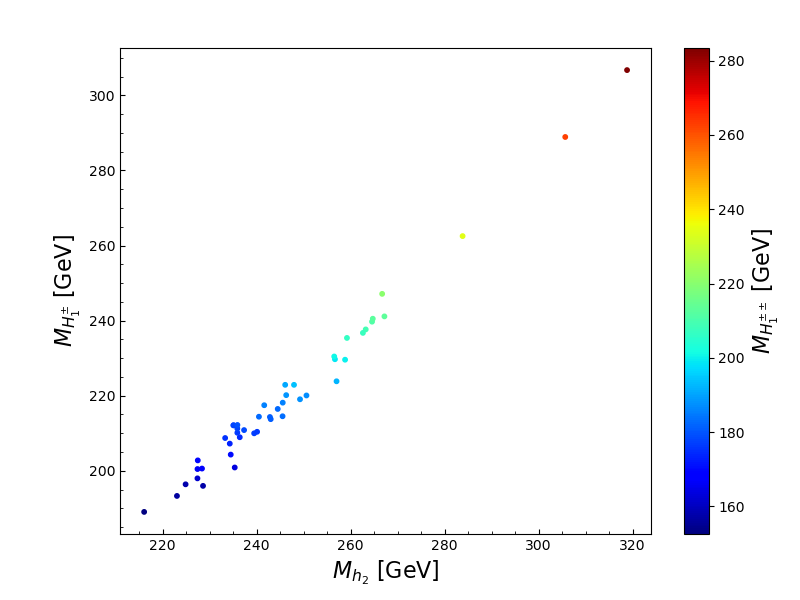}
\includegraphics[width=.32\textwidth]{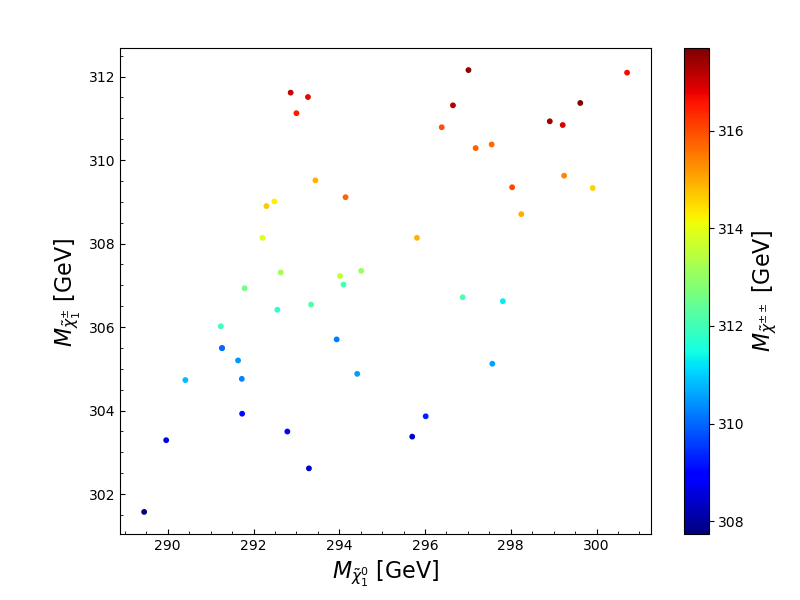}\\
\includegraphics[width=.32\textwidth]{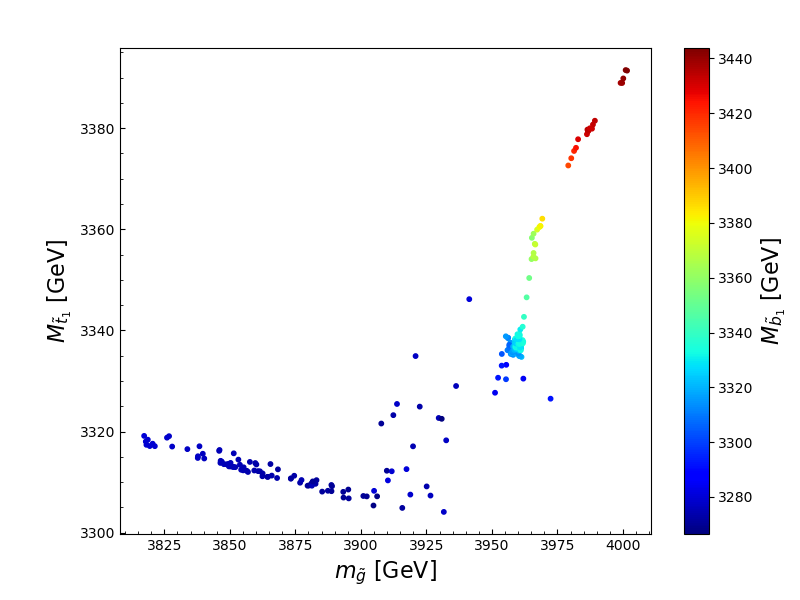}
\includegraphics[width=.32\textwidth]{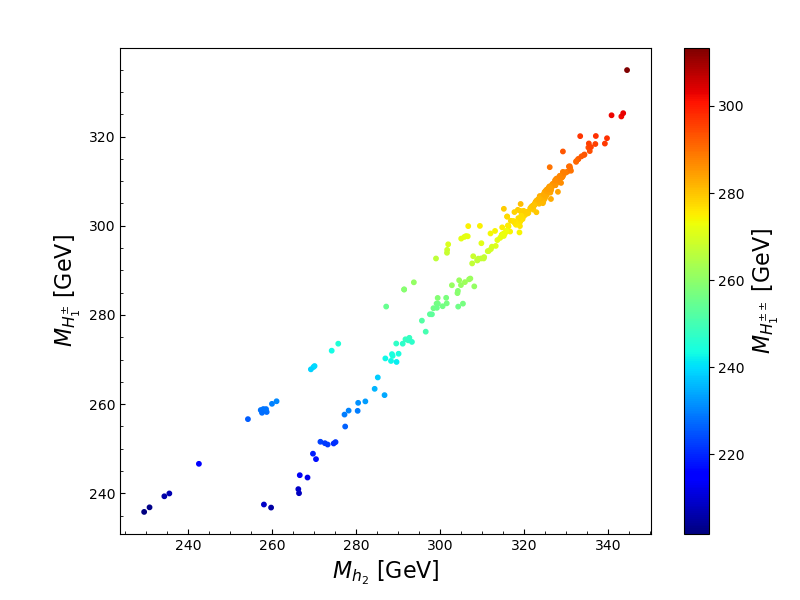}
\includegraphics[width=.32\textwidth]{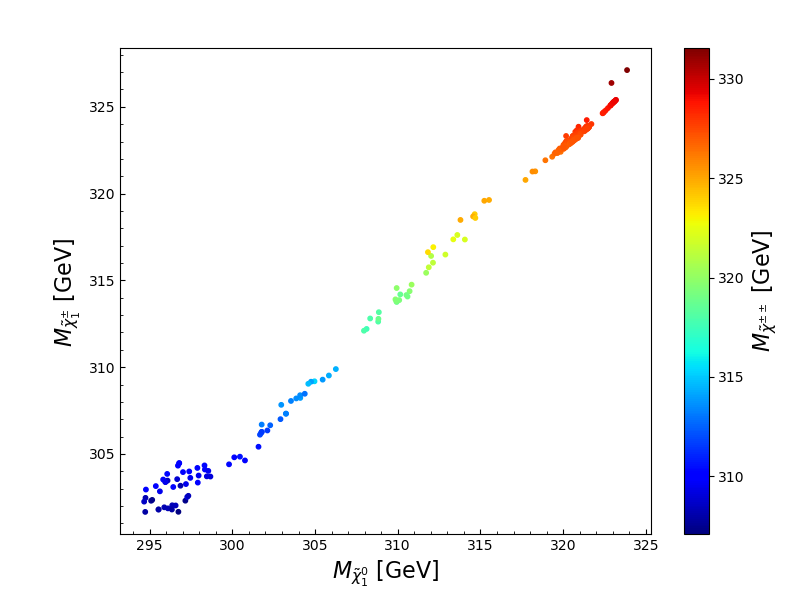}
\caption{The masses of typical particles in the SCTM. Left column: masses of the gluino and light top squark, where the color bar represents the mass of the light sbottom squark. Middle column: we present the masses of the next-to-lightest $CP$-even Higgs boson and the lightest charged Higgs boson, with different colors corresponding to the mass of the lightest doubly charged Higgs boson.
The masses of the lightest neutralino, the lightest chargino and the lightest doubly charged chargino can be found in the right column. 
}
\label{fig2}
\end{figure}

\begin{figure}
\centering
\includegraphics[width=.24\textwidth]{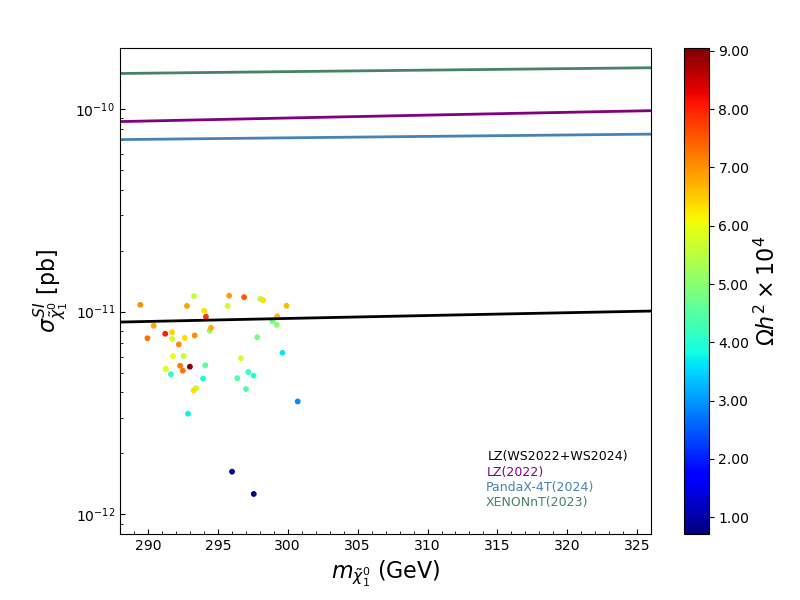}
\includegraphics[width=.24\textwidth]{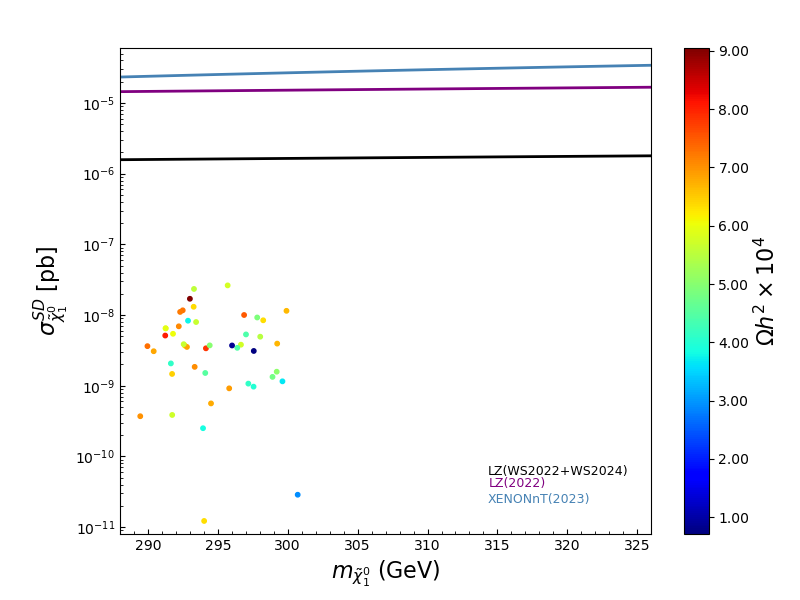}
\includegraphics[width=.24\textwidth]{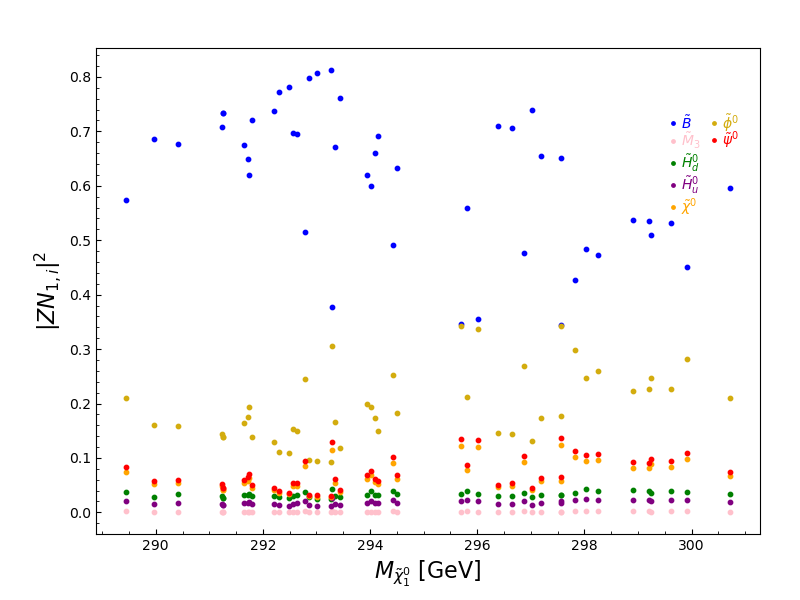}
\includegraphics[width=.24\textwidth]{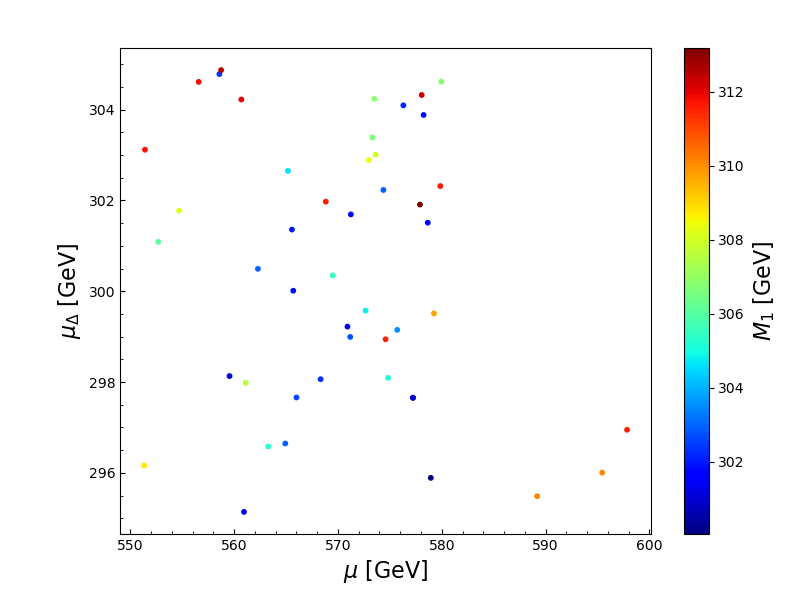}\\
\includegraphics[width=.24\textwidth]{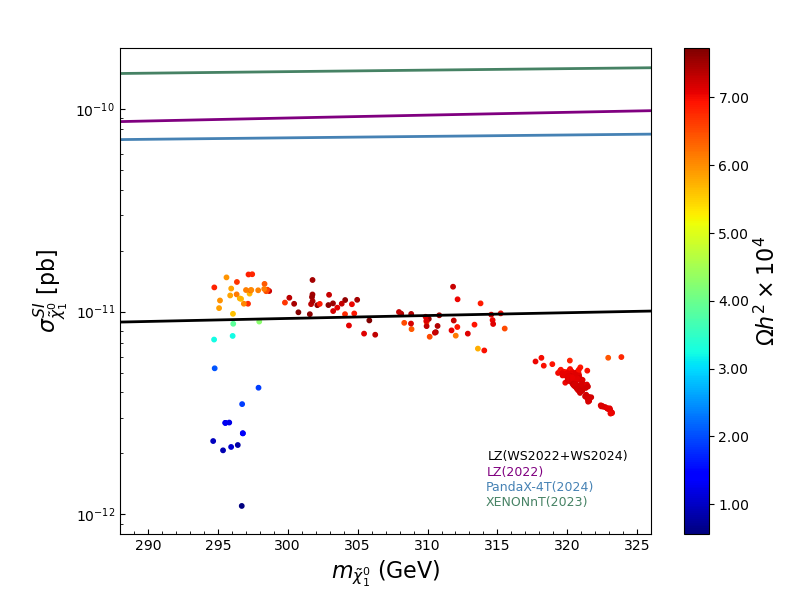}
\includegraphics[width=.24\textwidth]{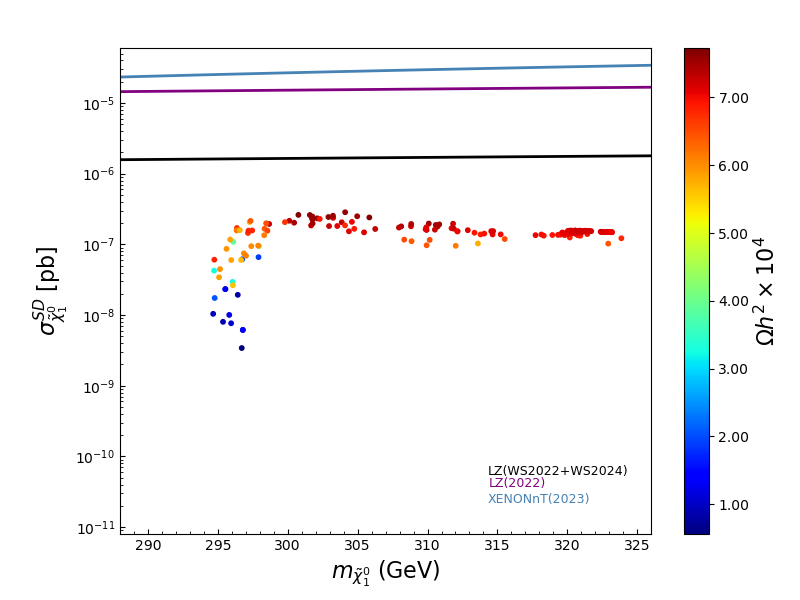}
\includegraphics[width=.24\textwidth]{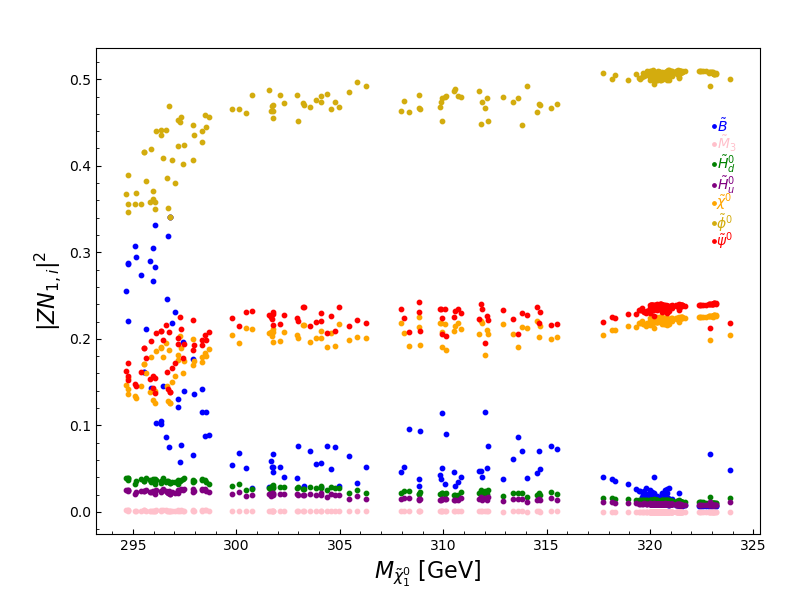}
\includegraphics[width=.24\textwidth]{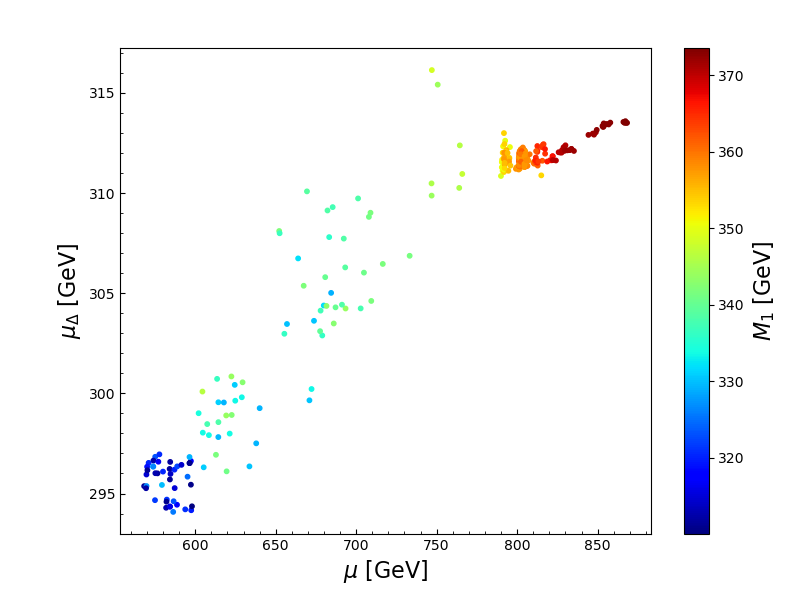}
\caption{The constraints from DM direct detection and the components of the DM candidate. The left two columns present the constraints of DM direct detection based on the spin-independent and spin-dependent interactions, respectively. 
The contribution of specific components for DM candidate can be found in the middle right column. In the right column, we show the typical mass term of the Higgsino, tripletino and bino.}
\label{fig3}
\end{figure}

It is not trivial to find the surviving parameter space that can provide a proper tripletinolike DM candidate. 
Here, all the samples shown in this section  satisfy the constraints presented in the previous subsection. Almost all of them successfully offer tripletinolike DM candidates, while a few provide only binolike DM candidates. Therefore, we  present the results of the binolike DM scenario in the upper row and the results of the tripletinolike DM scenario in the lower row for Figs. \ref{fig1}-\ref{fig3} in the following discussion.

The SM-like Higgs boson mass for both scenarios is well around 125 GeV as shown in the left column of Fig.~\ref{fig1}. The different colors correspond to the value of $\la$ and it is obvious that there exist two different parameter spaces in these samples. 
The relationship of the masses of light top squarks and  trilinear coupling $A_t$ with  the SM-like Higgs boson mass can be found in the middle column of Fig.~\ref{fig1}. We can see that both of these are quite large, which guarantees a considerable contribution to the SM-like Higgs boson mass. 
In the right column of Fig.~\ref{fig1}, we can tell the specific values of each component ($H_u^{0,r}$ in red, $H_u^{0,r}$ in blue, the triplets $\chi^{0,r},~\phi^{0,r}$ and $\psi^{0,r}$ in green, purple and orange, respectively) in the SM-like Higgs boson wave function. It is easy to see that  $H_u^{0,r}$ is dominant, which ensures one can fulfill the behavior of the 125 GeV Higgs boson as SM-like. The contribution of triplets can reach several percent, significantly impacting the mass of the SM-like Higgs boson and potentially contributing up to several GeV.

It is interesting to note that, in the tripletinolike DM scenario, the mass of the SM-like Higgs boson indeed increases slightly with the increase of the triplet VEVs. Moreover, it is evident that a larger trilinear coupling  results in a heavier SM-like Higgs boson, as expected. However, what makes this situation more intriguing is the seemingly counterintuitive observation that a heavier top squark corresponds to a lighter SM-like Higgs boson. This unexpected behavior can be attributed to the significant role played by the triplets as shown in the lower left panel of Fig.~\ref{fig1}.

The masses of colored sparticles in our work are quite heavy, as shown in the left column of Fig.~\ref{fig2}: the light top squark and light bottom squark are both around 3.3 TeV and the corresponding gluino is as heavy as 3.8 TeV. This spectrum is  free from   current constraints from the LHC on sparticles,  while it is unfortunate that these sparticles with color still cannot be tested at the High-Luminosity LHC 
(HL-LHC)~\cite{ZurbanoFernandez:2020cco}, so one has to wait for the next generation hadron colliders (such as the SPPC~\cite{CEPCStudyGroup:2018rmc} and FCC-hh~\cite{FCC:2018vvp}) to investigate this parameter space. The next-to-lightest $CP$-even Higgs boson, the lightest charged Higgs boson and the lightest doubly charged Higgs boson primarily originate from a $CP$-even quintuplet Higgs fields, with similar masses around 200 GeV for both scenarios, as shown in the middle column of Fig.~\ref{fig2}. Because the quintuplet has no interaction with fermions, it can only be produced mainly by vector boson fusion and vector boson associated production processes, which are significantly suppressed with respect to  gluon fusion. The production
cross section is only around $200$ {fb} for quintuplet states with mass of 200--300 GeV~\cite{SCTM10}. The mass of the next-to-lightest quintuplet state is about 500--600 GeV, whose production cross section is lower to $50$ {fb}~\cite{SCTM10}.  The lightest triplet in our work is also as heavy as 500--600 GeV, where the production cross section can reach $200$ {fb} as  it can be produced via gluon fusion~\cite{SCTM10}. All of the cross sections are generally too low to produce significant signals, hence, our samples satisfy the constraints from HiggsBounds~\cite{HiggsBounds1,HiggsBounds2,HiggsBounds3,HiggsBounds4,HiggsTools}, with the exception of the lightest Higgs states. 
We also consider charged-current Drell-Yan pair production of the custodial Higgs bosons at the lower end of the mass range we target. As discussed in \cite{Fermiophobic at LHC}, the $95\%$ exclusion limits from $W^+W^-$ and $ZZ$ searches by CMS \cite{CMS:2015hra} indicate that the cross section time BR rate of this process in the $H^0$ decay first channel, $\sigma(p p \to H^0 H^\pm) \times {\rm BR}({H^0\to W^+W^-})$, is less than approximately 500 fb when the mass of the heavy neutral Higgs is 250 GeV. For the second decay channel, with ${\rm BR}({H^0\to ZZ})$ used instead, the rate is less than approximately 100 fb. We have checked one of our Benchmark Point (BP) with $m_{h_2} = 229.58~\rm{GeV}$ and $ m_{H_1^{\pm}} = 235.83~\rm{GeV}$, 
where $h_2$ (playing the role of the $H^0$ above) is the second-lightest $CP$-even Higgs boson and $H_1^{\pm}$ (playing the role of $H^\pm$ above) is the lightest charged Higgs state, while the masses of the other Higgs bosons are above 600 GeV. We find $\sigma(pp \to h_2 H^{\pm}_1) = 3.416$ fb, which is much smaller than the corresponding CMS limit.

The masses of the lightest ewinos, as shown in the right column of Fig.~\ref{fig2}, are highly degenerate around 300 GeV. Moreover, the masses of other tripletinolike and binolike ewinos are also clustered around 300 GeV, as both typical mass terms, $\mu_\De$ and $M_1$, are approximately 300 GeV, as shown in the right column of Fig.~\ref{fig3}. This highly degenerate spectrum ewinos is the key to survive the current stringent bounds from the LHC.

It's known that the most stringent constraints for WIMP DM are from  direct detection experiments, such as LZ~\cite{LZ}, PandaX-4T~\cite{PandaX-4T} and XENONnT~\cite{XENONnT}, especially, the  most recently reported results by LZ~\cite{LZ} on the search for WIMP-like DM, which have pushed the spin-independent cross section of WIMP to nucleon to $\mathcal{O}(10^{-12})$ pb for  masses of $\mathcal{O}(50)$ GeV and lowered the spin-dependent cross section to $\mathcal{O}(10^{-7})$ pb in the same mass region. In this work, all the samples of binolike WIMP DM candidate scenario can satisfy the spin-dependent cross section constraints and most of these can also fulfill the spin-independent cross section constraints as shown in the upper row of Fig.~\ref{fig3}. Generally speaking, the binolike DM always provides too large relic density though it can survive the stringent direct detection limits. We can thus tell that there is quite a large tripletino component increasing the DM couplings to nucleons. Considering the suppressed interaction between the tripletino and $Z$ boson because of the custodial symmetry at tree level, the most important reason why we can get  small relic density is the combination of two mechanisms of the quintuplet Higgs funnel, which has been investigated thoroughly in~\cite{SCTM3}, and the coannihilation with tripletinolike ewinos. The typical mass terms of both tripletinos and bino are around 300 GeV as shown in the upper right panel of Fig.~\ref{fig3}, which ensures a highly degenerate mass spectrum, thereby significantly reducing the relic density. 

One key goal of this work is to find the parameter space providing tripletinolike WIMP DM candidate successfully. Indeed, we identify a part of parameter space that can fulfill these stringent DM direct detection constraints shown in the lower row of Fig.~\ref{fig3}. Here, it can clearly be seen from the lower right two panels that the typical mass term of tripletinos is smaller than the bino one and that the lightest neutralino is truly dominated by the tripletino. The spin-dependent DM direct detection constraints can  easily be satisfied for this scenario thanks to the suppressed interaction between the tripletino and  $Z$ boson because of the custodial symmetry at tree level, as already mentioned, while significant parameter space is excluded by the spin-independent DM direct detection constraints. 
Furthermore, similar to the binolike DM scenario, the key mechanisms why the relic density can be so small are the quintuplet Higgs funnel and the coannihilation with other tripletinolike ewinos plus binolike neutralino.

In summary,  although both of the highlighted scenarios cannot be checked in the current DM direct detection experiments,  all the surviving data samples provided herein will be testable  at  the forthcoming next generation of DM direct detection experiments, especially PandaX-nT. Furthermore, it might be possible to investigate the whole parameter space if these are combined with the results from  next generation hadron colliders.

\section{\label{sec-5} Conclusions}
\label{sec-5}
The SCTM,  a supersymmetric generalization of the Georgi-Machacek model,  predicts a complex mass spectrum and rich phenomenology, with sparticles like gluinos and top squarks having masses in the multi-TeV range, which are then not constrained by current LHC results (and only partially by the upcoming HL-LHC) but can only be so by the next generation of hadron colliders (like the SPPC and FCC-hh). In contrast, the rather degenerate Higgs spectrum that we have found can well be checked already at the planned HL-LHC.

Specifically, our work successfully identified viable parameter space regions for tripletinolike WIMP-type DM, a novel candidate with respect to traditional SUSY scenarios, both minimal and nonminimal \cite{Moretti:2019ulc}, despite the increasingly stringent constraints coming from DM direct detection experiments such as LZ, PandaX-4T and XENONnT. The suppressed gauge interactions of such a DM due to a built-in custodial symmetry and the significant coannihilation between tripletinolike ewinos and binolike neutralinos allow much SCTM parameter space to remain viable. Although some large portions of it are excluded by DM direct detection constraints, others are still promising and will be tested in a variety of future DM experiments.

\begin{acknowledgments}
This work is supported by the National Natural Science Foundation of China (NSFC) No. 12447167, by the Joint Fund of Henan Province Science and Technology R$\&$D Program No. 225200810092 and No. 225200810030, by the Startup Research Fund of Henan Academy of Sciences No. 231820011, by the Basic Research Fund of Henan Academy of Sciences No. 240620006, by the Graduate Innovation Fund of Henan Academy of Sciences No. 243320031, and by the Natural Science Foundation of Henan Province under Grant No. 232300421217. This work is also supported in part by the NExT Institute and the STFC Consolidated Grant No. ST/X000583/1.
\end{acknowledgments}


\end{document}